\documentclass[reprint,showpacs,footinbib,citeautoscript,floatfix]{revtex4-1}

\usepackage{graphicx}
\usepackage{amsmath}
\usepackage{amssymb}
\usepackage{mathrsfs}
\usepackage{bm}
\usepackage{array}
\usepackage{natbib}
\usepackage{soul}
\usepackage{natbib}

\newcolumntype {s}[1]{@{\hspace{#1}}} 
\usepackage{color}
\usepackage[usenames,dvipsnames]{xcolor}
\usepackage{wrapfig}
\usepackage[normalem]{ulem}

\begin{document}

\title{Unconventional superconductivity and quantum criticality in SmN$_{1-\delta}$}

\author{W.~F.~Holmes-Hewett$^1$, K.~Van~Koughnet$^1$, J.~D.~Miller, B.~J.~Ruck$^2$, H.~J.~Trodahl$^3$ and R.~G.~Buckley$^2$}

\affiliation{$^1$The MacDiarmid Institute
for Advanced Materials and Nanotechnology and The School of Chemical and Physical Sciences, Victoria University of Wellington,
PO Box 600, Wellington 6140, New Zealand,}

\affiliation{$^{2}$The MacDiarmid Institute
for Advanced Materials and Nanotechnology and Robinson Research Institute, Victoria University of Wellington,
PO Box 600, Wellington 6140, New Zealand,}

\affiliation{$^3$The School of Chemical and Physical Sciences, Victoria University of Wellington,
PO Box 600, Wellington 6140, New Zealand}

\date{\today}


\begin{abstract}

Nitrogen vacancy doped SmN$_{1-\delta}$ is a semiconductor which lies in the intermediary between insulating-ferromagnetic SmN and metallic-anti-ferromagnetic Sm. The dopant electrons resulting from nitrogen vacancies have recently been predicted to lie in a band precipitated by a majority-spin 4\textit{f} level on the six Sm ions neighbouring each nitrogen vacancy, an in-gap state $\sim$1 eV below the 4\textit{f} states in stoichiometric SmN. Optical data reported here corroborate the prediction along with an extended computational study. Electrical transport measurements show the transition from an insulating to metallic state with a hopping type conductivity in dilutely doped films. We provide strong evidence that electron transport is mediated by a dispersion-less majority spin defect band implying triplet type superconductivity, the location of which in the SmN-Sm phase diagram suggests the location of a quantum critical point.


\end{abstract}

\maketitle


\section{Introduction}


The borders of ordered magnetic states in $f$-electron systems are a common source of emergent phenomena, including unconventional superconductivity and heavy Fermion behaviour~\cite{Wirth2016,Pfleiderer2009,Basov2011,Coleman2007,adler2019,Taillefer2010,Dzero2016}. Significant effort has been placed into understanding these phenomena in terms of zero temperature phase transitions and their accosiated quantum critical points (QCP), as these can act as the source of the magnetic instabilities which drive these exotic states~\cite{Kirchner2020,Komijani2019,Coleman2001,Keimer2017}. Commonly QCPs lie at boundaries between antiferromagnetic (AFM) and paramagnetic (PM) phases. However there is recent interest in ferromagnetic-phase boundaries~\cite{Brando2016,Rai2019,Hafner2019,Shen2020}. Lanthanide compounds have led as examples, and although the rare earth mononitrides (\textit{L}N, \textit{L} a lanthanide) were proposed already some years ago as likely heavy-Fermion candidates~ \cite{Wachter2001,Petit2010,Petit2016}, that prediction has not yet been explored in great detail.

 \begin{figure}
\centering
\includegraphics[width=\linewidth]{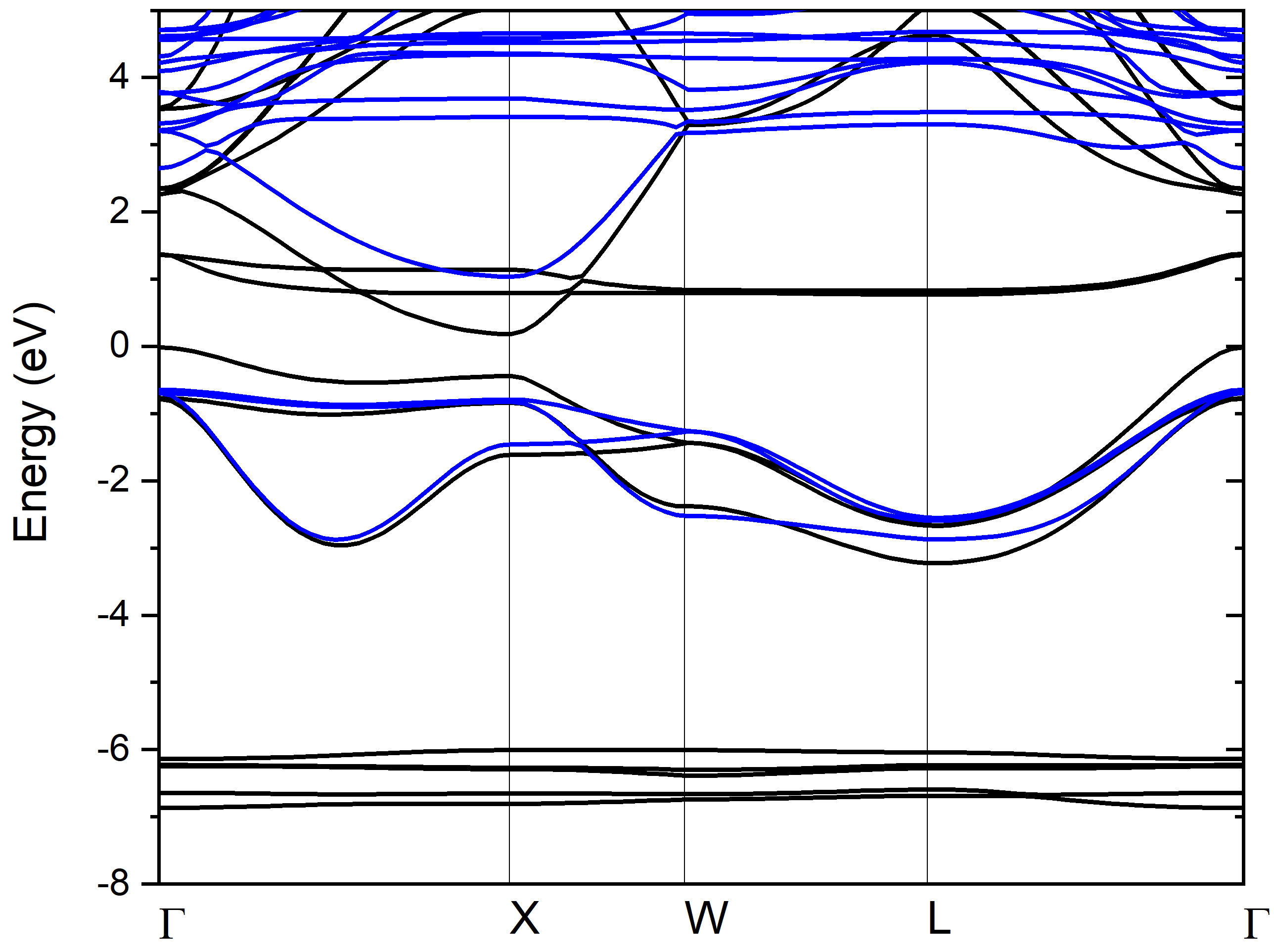}
\caption{Calculated band structure of SmN based on the primitive unit cell. Majority and minority spin states are in black and blue respectively.}
\label{Pristine}
\end{figure}


The \textit{L}N (\textit{L}$^{3+}$N$^{3-}$) are poised at a metal-insulator boundary, and ongoing studies over the last half century have placed them variously on the two sides of that boundary. Recent thin films are mostly on the insulator side, showing  signatures of a dopable semiconducting ground state with a narrow but clear band gap~\cite{Trodahl2007,Mitra2008,Yoshitomi2011,Vidyasagar2014}. Electron doping in thin films can be routinely controlled by a residual concentration V$_N$ of nitrogen vacancies~\cite{Punya2011,Plank2011,Holmes-Hewett2020,Devese2022}. They crystallise into the NaCl structure with lattice constants varying from 0.51~nm to 0.48~nm across the series~\cite{Natali2013}. Within that structure the cations form a close-packed FCC network that differs only in the stacking sequence from that in the hexagonal structures of the pure lanthanide metals. Remarkably, in addition to their similar local close-packed arrangements, the \textit{L}-\textit{L} separation is only slightly different; in particular the Sm-Sm separation in SmN is 0.3560 nm vs 0.3606 nm in metallic Sm, a contrast of only 1.3\%.

The picture that emerges is then of close-packed lanthanide lattice with nitrogen ions entering the network with minimal influence on the \textit{L} packing density. They each remove three electrons from the \textit{L}~5$d$ and 6$s$ states that form the conduction channel, finally reducing the mobile electron concentration to zero in stoichiometric \textit{L}N. In the pure metallic \textit{L} phase the magnetic exchange, ferromagnetic for most, is dominated by an RKKY interaction via those mobile electrons, and nesting across portions of the Fermi surface then lead to the rich range of spiral spin alignments revealed by neutron scattering studies fifty years ago. In contrast the nitrides, also mostly ferromagnetic, involve indirect exchange via the \textit{L}~5\textit{d} and N~2\textit{p} states~\cite{Sharma2010,Larson2006}.

Interest in the \textit{L}N series has thrived for well over half a century, although a focus on their magnetic properties has been only rarely extended to discussions of their strong correlation~\cite{Wachter2001,Petit2010,Petit2016}. In general most attention has been expended on GdN, with the series' highest Curie temperature of $\sim$~65~K. The half-filled 4\textit{f} level, configuration $^8$S$_{7/2}$ dictates that there is no orbital contribution to the magnetic moment; it is an entirely conventional spin-only ferromagnetic compound. Unlike GdN, other \textit{L}N can feature dopable strongly correlated 4\textit{$f$} states~\cite{Holmes-Hewett2018,Holmes-Hewett2019a,Degiorgi1990,Holmes-Hewett2021} which promise new emergent behaviour. In that context, there are reports of superconductivity in SmN~\cite{Anton2016} and the Kondo effect in EuN~\cite{Binh2013}. 

\begin{figure*}
\centering
\includegraphics[width=\linewidth]{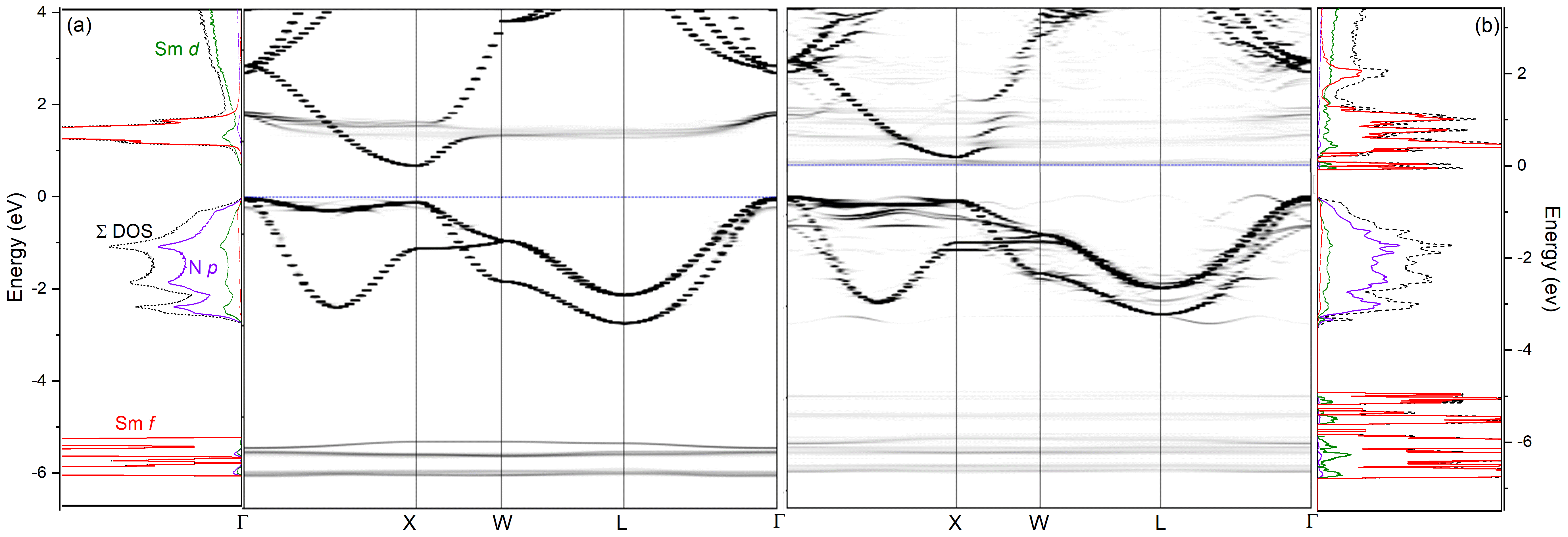}
\caption{Calculated band structure and DOS for undoped (a) and nitrogen vacancy doped (b) SmN supercells. The shading on the in-gap defect states at the Fermi energy in (b) has been darkened for clarity.}
\label{bands}
\end{figure*}

The Sm$^{3+}$ ion in SmN has five 4\textit{f} electrons, two fewer than Gd$^{3+}$, ensuring that there are two empty majority-spin 4$f$ bands threading the 5\textit{d} CB~\cite{Larson2007,Preston2007,Galler2022,Morari2015} as shown in Figure~\ref{Pristine}. The inter-ion exchange precipitates ferromagnetic spin alignment of those five occupied 4\textit{f} states below $\sim$~30~K. However, within the $^6$H$_{5/2}$ state the spin magnetic moment is opposed by an orbital moment of similar magnitude so that the \textit{net} magnetic moment is nearly zero; SmN displays ferromagnetic alignment and exchange split bands but has a near net zero moment of $\sim$~0.035~$\mu_B$ per Sm ion~\cite{Meyer2008,Anton2013,McNulty2016}. Electron doping in SmN is significantly more complex than in the \textit{L}N with 4\textit{f} bands removed from the Fermi energy. In SmN the three electrons released by a nitrogen vacancy do not appear in the intrinsic 5\textit{d} conduction band minimum~\cite{Holmes-Hewett2021}. They rather occupy majority-spin 4\textit{f} states on the six Sm ions neighbouring the vacancy. These states appear $\sim$~1 eV lower than the 4\textit{f} states on fully nitrogen coordinated ions, ultimately lying close to mid-gap and pinning the Fermi energy (see Figure~\ref{bands}(b)). These defect states hybridise with both the 5\textit{d} and, more weakly, with the N~2\textit{p}.

Within the LN series series the Sm/SmN pair stands out as the one in which the end points have contrasting magnetic order, ferromagnetic below 30~K in the mononitride but antiferromagnetic below 100~K in metallic Sm~\cite{Yao1980,Thompson1992}. It is also the only LN in which superconductivity has been reported across that range, in heavily doped (i.e. nitrogen deficient) SmN$_{1-\delta}$~\cite{Anton2016}, motivating a search for heavy-Fermion superconductivity that is commonly found  near a QCP~\cite{Mathur1998,Pfleiderer2009}. Here we report a combined experimental and computational study of the band structure and defect states that precipitate superconductivity in SmN.

 \begin{figure}
\centering
\includegraphics[width=\linewidth]{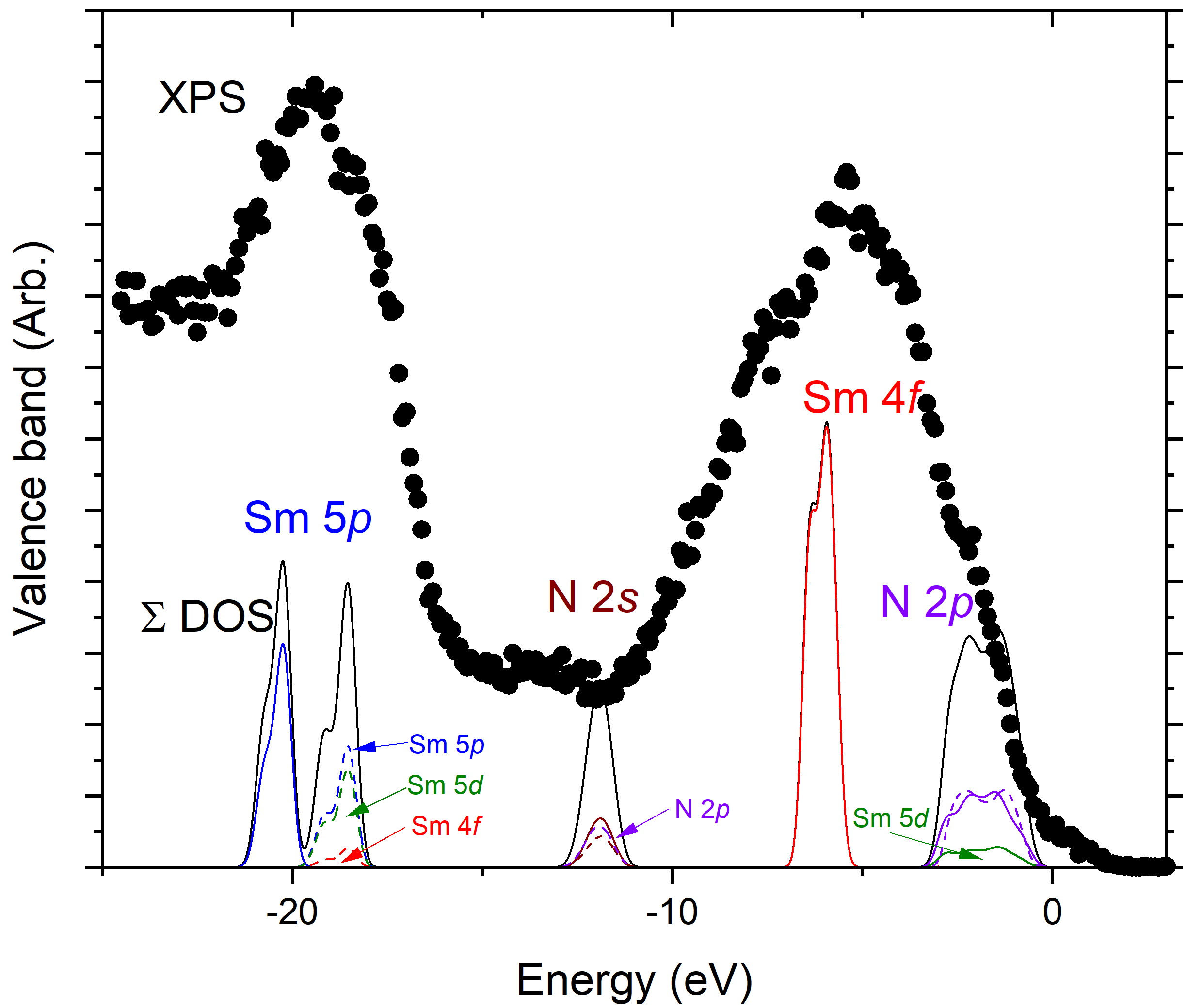}
\caption{XPS measurements (black circles) and DOS calculations (lines) for SmN. The total DOS (majority and minority spin) are shown in solid black lines, the contribution from separate orbitals are shown in various colours (majority spin - solid lines, minority spin - dashed lines), for clarity only the significant contributions to each peak are shown.}
\label{valenceband}
\end{figure}


\section{Experimental and computational Details}

SmN thin films were grown in various ultra-high vacuum chambers with base pressures on the order of 10$^{-9}$~mbar~(see reference~\cite{Natali2013} for details). A range of substrates have been used, (Al$_2$O$_3$, Si and SiO$_2$) selected for ease of electron transport and optical measurements. Once grown to the chosen thickness (roughly 100~nm) the films were passivated with insulating AlN. Sm was evaporated at a flux of $\sim$~1~$\mathrm{\mathring{A}}$/s in the presence of molecular nitrogen at varying pressures from $1~\times 10^{-6}$~mbar to $4~\times 10^{-4}$~mbar to control V$_N$~\cite{Ullstad2019}. As a further level of control the substrate temperature during growth was varied in the range of 300~K to 700~K. X-ray diffraction confirmed all films were rock-salt SmN. 

Electrical measurements were conducted in a Quantum Design Physical Properties Measurement System with Au contacts in a van der Pauw configuration. The optical transmission and reflection measurements were conducted at room temperature between energies of 0.01~eV to 4~eV in a Bruker Vertex 80v Fourier transform spectrometer. Reflection measurements were referenced using an Al film, and the results then adjusted for the finite reflectivity of Al~\cite{Ehrenreich1963}. The optical measurements were modelled using the software package RefFit~\cite{RefFit}.

The X-ray photo-emission spectroscopy (XPS) measurements were performed using a Kratos XSAM 800 spectrometer. An Al source was used to provide monochromated K alpha x-rays. During analysis, the operating pressure was typically $8\times10^{-9}$~mbar or better. To remove the AlN passivation layer, the samples were sputtered using Ar$^+$ ions.

Density functional theory based calculations were undertaken using the \emph{pwscf} family of code of Quantum Espresso~\cite{QE,Cococcioni2005,Topsakal2014}. Self-consistent calculations on the primitive cell were completed using a $k$ mesh with $10\times10\times10$ divisions, while super-cell calculations were on a $4\times4\times4$ division \textit{k}-mesh. The wave function and charge density cut-off energies were 50 Ry and 200 Ry respectively for all calculations. 

The 4$f$ electrons of the LN series are strongly correlated and thus require careful treatment beyond the traditional DFT methods~\cite{Larson2006,Larson2007}. In the basic DFT the 4$f$ states would be found at the Fermi energy for most of the LN. In reality the strongly correlated nature of these bands pushes the filled states below and unfilled states above the Fermi energy. This physics can be approximated using the DFT+$U$ method where the behaviour of the correlated orbitals is determined by an adjustable parameter $U$. In the present study two $U$ parameters are used, as described first in reference~\cite{Larson2006}. One to account for the strongly correlated 4$f$ states ($U_f$), and a second applied to the 5$d$ states ($U_d$). The latter is used to correct for the underestimate of the optical band-gap, which is known for the LSDA method. For details see our previous publication~\cite{Holmes-Hewett2021}.



Following conventional DFT+$U$ calculations the output from Quantum Espresso was used to generate maximally localised Wannier functions using Wannier~90~\cite{Mostofi2014,Marzari1997,Souza2001}. The resulting Wannier functions were then used to calculate the density of states (DOS) and joint density of states (JDOS) on denser \textit{k}-meshes of $25\times25\times25$ divisions.

\section{Results}

\subsection{Computational studies}

We begin our results with a brief comparison of the calculated and experimental valence bands. Figure~\ref{valenceband} shows the experimental XPS spectrum along with the calculated DOS in the valence band of SmN. The XPS data show three separate features in the range from $\epsilon_f$ to $\epsilon_f-25$~eV. The lowest energy feature in the experimental data, with a centre near -19.5~eV, matches well with the calculated position of the feature dominated by the Sm~5\textit{p} states. The calculation shows two spin-split peaks, the lower being formed largely from the Sm~5\textit{p} majority spin states, while the higher is a mix of various minority states, Sm~5\textit{p}, 5\textit{d} and 4\textit{f}. The experiment is carried out at ambient temperature in the paramagnetic phase, so one would expect the majority and minority Sm~5\textit{p} states to be degenerate at intermediate energy. These then form the bulk of the feature. Moving to higher energy the N~2\textit{s} states are placed by the calculation at $\sim-12$~eV, however these are not clear in the measurement. There is an indication of a weak peak roughly 1~eV lower at -13~eV which could correspond to these states. Moving now to the valence band maximum (VBM) which can be seen to be formed from N~2\textit{p} states. The calculation finds these are hybridised with the majority spin Sm~5\textit{d}. Finally the peak in the XPS data at $\sim$~-5.25~eV matches well with the calculated location of the Sm~4\textit{f} states, of which majority spin only are present. The correspondence between the calculated Sm~4\textit{f} states and the experimental XPS spectrum is significant as this indicates the calculated value of $U_f=6.77$~eV is appropriate. The XPS data are much broader than the corresponding calculated DOS, particularly for the Sm~4\textit{f} states, as is also observed in ErN~\cite{Mckay2020}. This is largely due to limitations in the calculation where an effective Hubbard term $U_{eff}=U_f-J_f$ is used rather than the full treatment of $U_f$ and $J_f$, the use of the latter would encourage additional splitting between occupied 4\textit{f} states, as appears to be the case in the physical material.

The calculated band structure for crystalline SmN shown in Figure~\ref{Pristine} is based on the primitive unit cell. It is interesting to consider that under the octahedral symmetry of the lattice, the seven 4\textit{f} wave-functions for each spin are split into three groups of degeneracy 3, 3 and 1. The degeneracy of these groups dictate that the 4\textit{f}$^5$ configuration of the Sm$^{3+}$ ion must have a metallic ground state, with degenerate filled and unfilled 4\textit{f} levels meeting at the Fermi energy. This is clearly not the case as experimental results agree on a semiconducting, or insulating ground state when the material is undoped. The 4\textit{f} wavefunctions must then break the cubic symmetry, with the resulting band structure shown in Figure~\ref{Pristine}. We note that the 4\textit{f} wave functions not conforming to the lattice symmetry is well understood and contributes to the finite \textit{un}quenched orbital angular momentum in many rare-earth compounds. Under octahedral symmetry the N~2\textit{p} states have a three fold degeneracy at $\Gamma$ which is also broken in Figure~\ref{Pristine}. We see in Figure~\ref{Pristine} the $p_z$ states are $\sim$~1~eV higher than the $p_x$ and $p_y$ states. This results in the vanishingly small indirect $\Gamma$-X gap of $\sim$~0.15~eV. 

Any physical crystal will deviate somewhat from the pristine structure imposed by the periodic boundary conditions applied to the primitive unit cell. To computationally investigate something closer to the physical material we have relaxed an undoped 54 atom super-cell seeded with a small amount of disorder. The resulting band structure, unfolded to represent the familiar bands of the primitive unit cell, is shown in Figure~\ref{bands}(a) (for clarity the majority spin bands only are shown). The band structure shown in Figure~\ref{bands}(a) is indeed very similar to that of the pristine cell shown in Figure~\ref{Pristine}, with some notable differences. The most significant contrast is between the N~2\textit{p} bands at the VBM. In the relaxed crystal the three N~2\textit{p} states are roughly degenerate at $\Gamma$ near the energy of the $p_x, p_y$ bands in the pristine cell. The VBM is lowered, increasing the indirect gap to $\sim$~0.75~eV which is very slightly smaller than the optical gap at X of 0.8~eV. 

Figure~\ref{bands}(b) shows the unfolded bands of a SmN super-cell doped with a nitrogen vacancy at $\sim$3~\% concentration. As described in our earlier work~\cite{Holmes-Hewett2021}, the structural disorder resulting from the vacancy results in long range periodicity and thus new defect states in the crystal. These manifest as the \textit{ghost-like} bands, the shading of which signifies the weight of the unfolded state for a given \textit{k}-point. The most striking feature, and clearest difference from both the undoped cell and the other nitrogen vacancy doped LN, is that the Fermi energy resides in the intrinsic gap region rather than the VBM or the CB. It is pinned here by defect states which are largely localised to the six Sm ions which coordinate the vacancy site~\cite{Holmes-Hewett2021}. This is a clear contrast to LuN~\cite{Devese2022} and GdN~\cite{Punya2011}, the only other V$_N$ doped LN investigated computationally, where nitrogen vacancy doping lifts the Fermi energy into the L~\textit5{d} CB in a more conventional manner. The structural and charge disorder induced by the vacancy causes the 4\textit{f} states throughout the crystal to become more localised than in the pristine environment, to an even greater degree than the perturbed undoped cell in Figure~\ref{bands}(a). The \textit{bands} of the pristine cell are gone and replaced with the \textit{ghost-like} smear of more localised 4\textit{f} states which span the CBM.

\subsection{Experimental results}

\begin{figure}
\centering
\includegraphics[width=\linewidth]{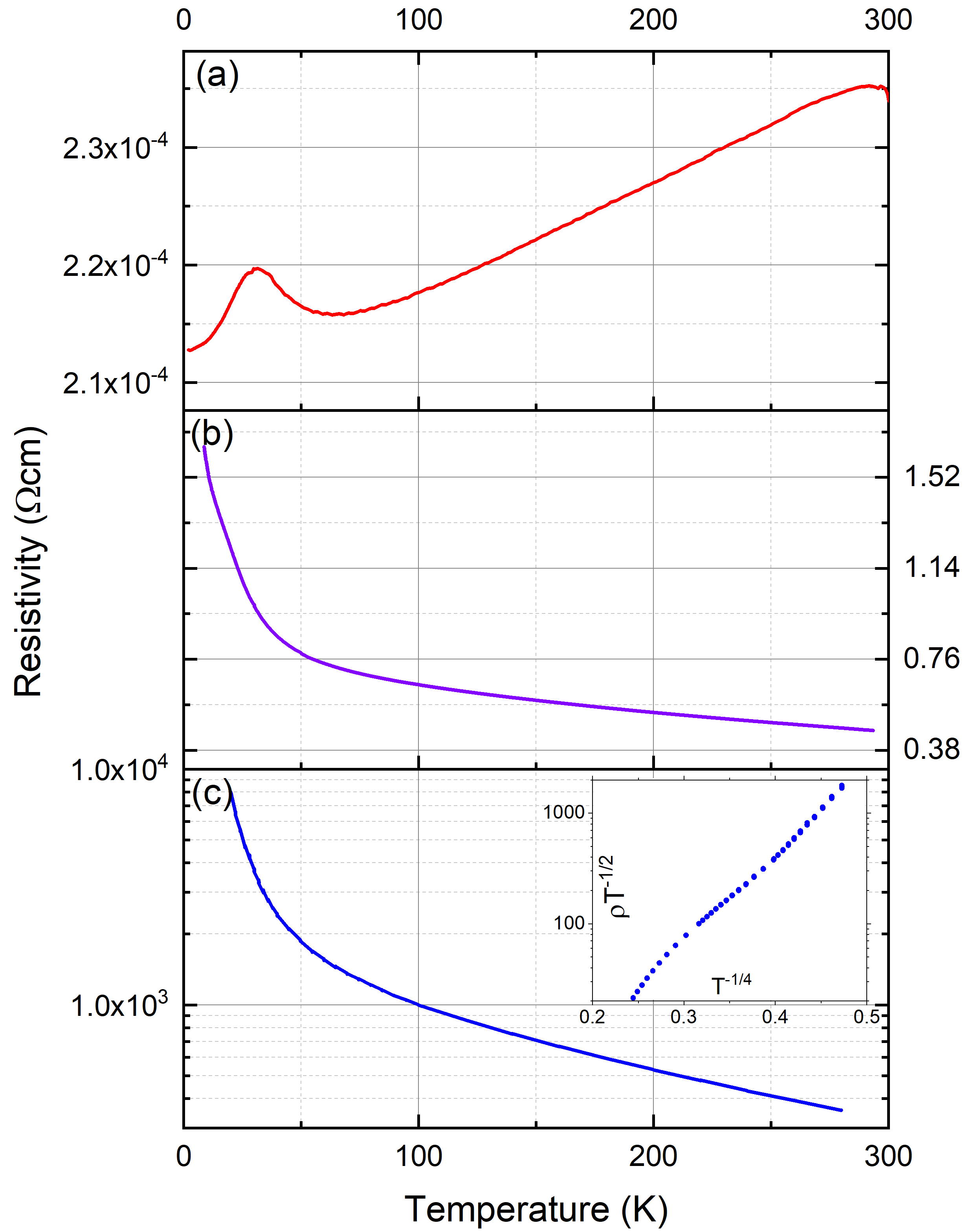}
\caption{Resistivity as a function of temperature for three SmN films produced to harbour significantly different concentrations of nitrogen vacancies. Measurements are shown in zero applied field (solid lines) and 3~T (dashed lines). Panel~(a) shows a film with a metallic like conductivity. Panels (b) and (c) show films with non-metallic conductivities.}
\label{SmN_Res}
\end{figure}

With the computational studies in mind we now turn to the experimental data. The runs of resistivity with temperature in Figure~\ref{SmN_Res} show clearly the contrasting behaviour between films grown with various concentrations of V$_N$. The most conductive film in panel~(a) shows a positive coefficient of temperature; it is clearly doped to degeneracy. The more nearly stoichiometric films show resistivities that diverge at the lowest temperatures. The most dilutely doped film has a resistivity which increases strongly with decreasing temperature, and at the lowest temperatures shows a temperature dependence ($\rho\sim \mathrm{exp}[(T_0/T)^{1/4}]$) characteristic of variable-range hopping. The anomaly near 20~K in the conductive film is close to the ferromagnetic transition. This is a result of magnetic disorder scattering in this film~\cite{Maity2018, DeGennes1958, Fisher1968}. A similar feature is present in our moderately doped films (not shown) and results from a band gap reduction across the Curie temperature~\cite{Trodahl2007}. This feature is obscured in the most resistive films by the rise in resistivity at low temperature.

\begin{figure}
\centering
\includegraphics[width=\linewidth]{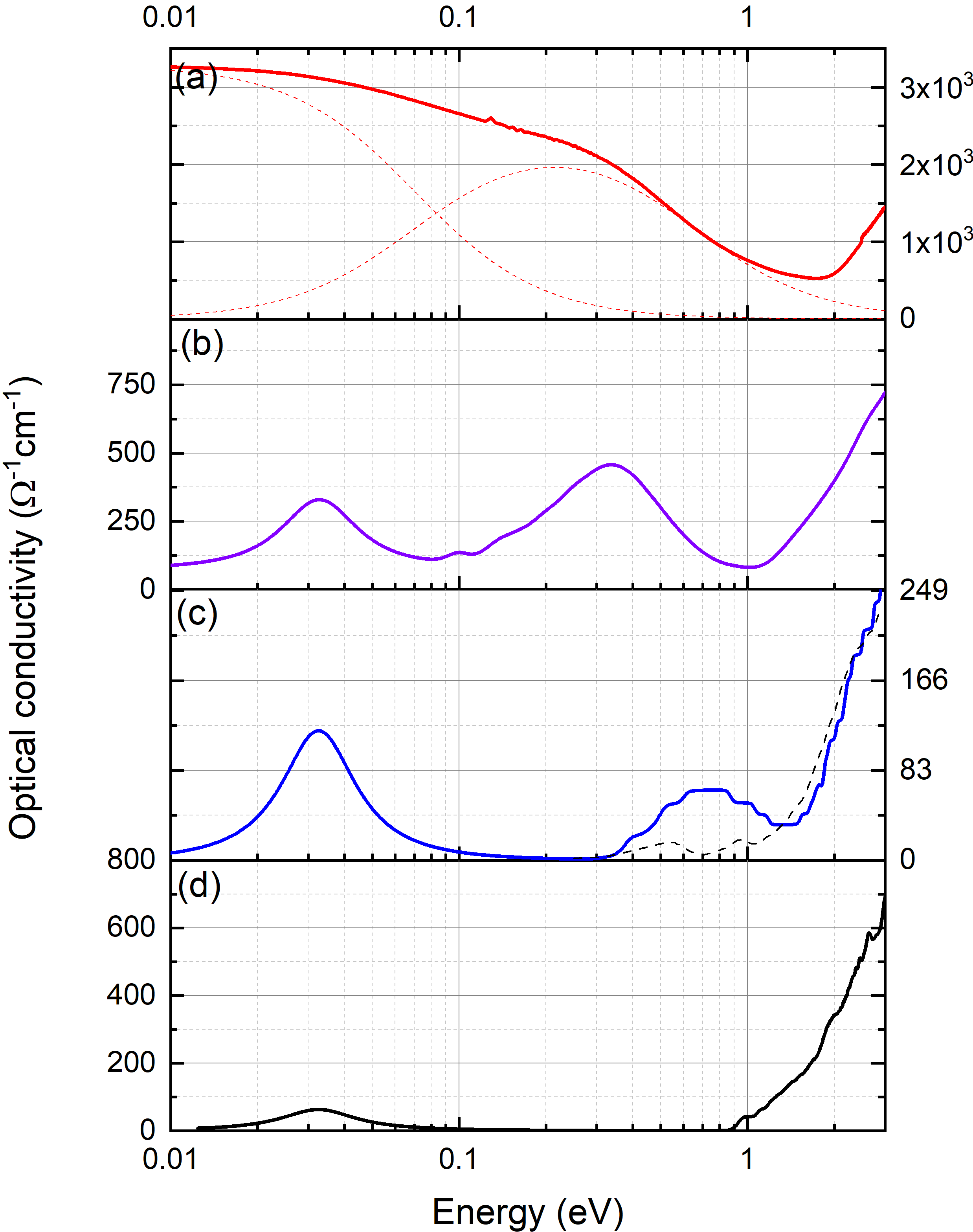}
\caption{Optical conductivity for four SmN films based on reflection and transmission measurements. Panel (c) shows in addition the joint density of states corresponding to the band structure of figure~\ref{bands}(b). The colours are consistent between figures~\ref{SmN_Res} and \ref{Optical} and represent measurements on the same films.}
\label{Optical}
\end{figure}

Far IR to near UV optical spectroscopy was performed on a series of films doped variously with nitrogen vacancies leading to dc (zero-frequency) conductivities ranging from $\sim$~1$\times$10$^{-4}~\Omega^{-1}$cm$^{-1}$ to $\sim$~5000~$\Omega^{-1}$cm$^{-1}$ measured at 2~K. Figure~\ref{Optical} shows the resulting optical responses of four films, two from the extremes of our conductivity and two representative of the centre. 

We begin with the most insulating film in panel (d) of Figure~\ref{Optical}. This film is the closest to stoichiometric that we have grown, and as such has a dc resistivity above our measurement limits. The optical conductivity is vanishing at low frequencies indicating any unintended nitrogen vacancies have little effect on the electrical transport. Moving to higher energy at $\sim$30~meV there is an absorption relating to the IR active TO~$\Gamma$ mode vibration which we see in all our LN films~\cite{Holmes-Hewett2022}. There is then little absorption in the MIR region before an increase near $\sim$~1~eV signalling the onset of optical transitions. Moving to the next most insulating film in panel (c) we now see a feature begin to develop below the intrinsic optical gap, as discussed in our previous report~\cite{Holmes-Hewett2019}. Panel~(c) also shows the JDOS corresponding to the $\sim$~3~\% electronic structure of the material illustrated in Figure~\ref{bands}(b). The JDOS shows a similar form to the optical conductivity with a double peaked MIR feature corresponding to transitions into and out of the in-gap states, before the high energy rise corresponding to the intrinsic optical gap. The relative strength of the MIR feature is weaker in the JDOS than in the optical conductivity, this may indicate the matrix elements coupling the initial and final states are enhanced when these transitions involve defect states over the VB to CB transitions of the stoichiometric material.  

The moderately conductive film in panel (b) now has a finite zero-frequency conductivity consistent with the electrical measurements in Figure~\ref{SmN_Res}. The MIR feature has grown in magnitude and shifted to lower energies. Finally the most conductive film in panel (a) has a strong free carrier absorption at the lowest energies, which matches well the measured dc conductivity. The MIR feature has now softened with a centre near 0.2~eV.

\begin{figure}
\centering
\includegraphics[width=\linewidth]{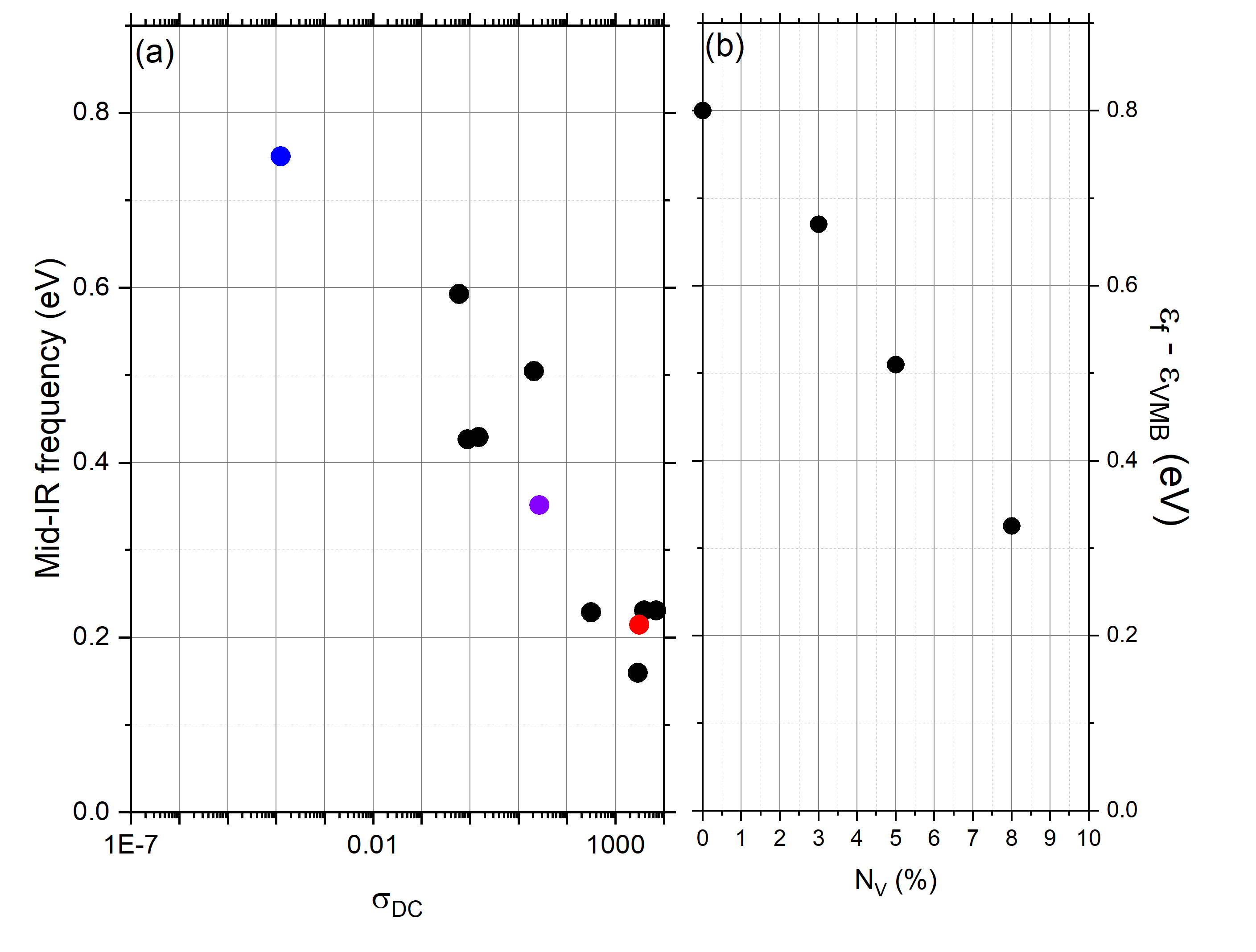}
\caption{(a): centre of the mid infra-red feature for a series of SmN films, the coloured points indicated films in Figures~\ref{SmN_Res} and \ref{Optical}. (b): difference between the Fermi energy and valence band maximum for calculations over a range of nitrogen vacancy doping concentrations.}
\label{Optical2}
\end{figure}

The development of the mid-IR absorption is the most important change with V$_N$ across Figures~\ref{Optical} (a) to (d), along with its softening to lower energy as the conductivity increases. The strength of the feature appears also to increase significantly in the most heavily doped film (Figure \ref{Optical}(a)), though one must note that the amplitude of the feature is more susceptible than the red shift to interference from the fitted zero-energy form used to represent the free-carrier conductivity. Lorenztian fits to the feature in the full set of films yield the central energy plotted versus the conductivity in Figure~\ref{Optical2}(a). This shows a sharply reduced energy as the conductivity rises, falling from $\sim$1~eV in near-stoichiometric film to $\sim$0.2~eV in heavily doped films.

In order to understand the softening of the mid-IR feature, we turn to the calculated band structure in N-deficient SmN~\cite{Holmes-Hewett2021} and Figure~\ref{bands}(b) in the present manuscript. As noted above a defect band dominated by the 4\textit{f}-states on the Sm ions neighbouring a vacancy forms within the SmN fundamental gap when the crystals are doped with V$_N$. The Fermi level is then pinned to these in-gap defect states, forming an increasingly extended-state band as the N-vacancy concentration, and thus doping, increases. The calculations show that with an increasing V$_N$ the in-gap N-vacancy states move deeper into the gap effectively pulling the Fermi energy towards the valence band maximum. The defect band is strongly localised with weight at all wave vectors, which is then effectively free of the wave-vector conservation selection rule. As these states span the Fermi energy they can harbour both final and initial states for an optical transition involving any of the extended-state bands in the CB and VB. On that basis we look at transitions involving the mid-gap 4\textit{f} states, which is effectively a measurement of the separation between the VBM at $\Gamma$ and the Fermi energy on one side, and the separation between the Fermi energy and the unoccupied states at the CBM on the other side. In Figure~\ref{Optical2}(b) we have plotted the separation in energy between the valence band maximum and Fermi energy of four concentrations examined computationally (Note that for the 0~\% doping case the value represents the minimum optical gap between the VBM and CBM). The agreement between the optical data and the calculation are striking and suggest that the transitions from the VBM to the defect states at the Fermi energy are the dominant contribution to the MIR feature in the optical spectra. The clarifies our previous report~\cite{Holmes-Hewett2019} where we interpreted the MIR feature as locating the unfilled majority spin 4\textit{f} in SmN. The present study shows the situation is more complex. The MIR feature clearly tracks the defect states which are accommodated by Sm ions adjacent to the vacancy site, rather than the intrinsic unfilled majority spin 4\textit{f} states.

\begin{figure}
\centering
\includegraphics[width=\linewidth]{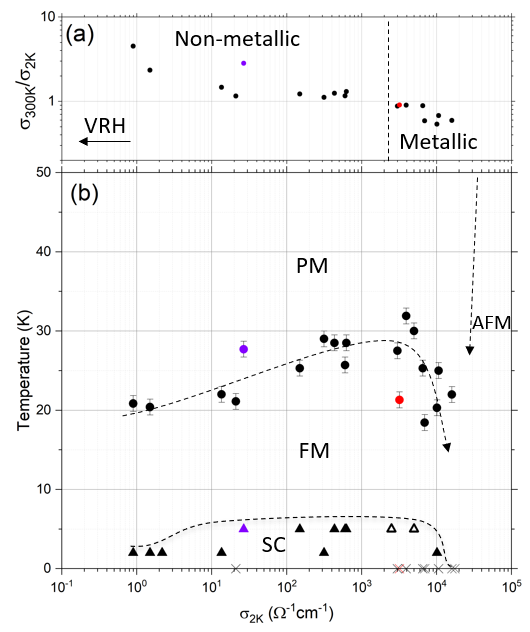}
\caption{(a): Ratio of 300~K conductivity to 2~K conductivity for the series of SmN films. (b) Phase diagram for the series of SmN films showing the paramagnetic, ferromagnetic and superconducting phases of SmN and anti-ferromagnetic phase of metallic Sm.}
\label{Phase_D}
\end{figure} 

These dynamic defect states have interesting implications regarding electron transport. Contrasting the other LN, electrical transport in SmN appears to be mediated by the in-gap states rather than extended states in the intrinsic CB. In the dilute doping case the occupied in-gap states are localised, separated in real-space from one-anther. This then results in the variable range hopping type conductivity apparent in the dilutely doped film of figure~\ref{SmN_Res}(c). Eventually, as the concentration of V$_N$ increases, a percolation limit is found, along with a metallic-like ground state. This metallic state is characterised by the small resistivity seen the heavily film in figure~\ref{SmN_Res}(a) and the conventional Drude like roll-off, observed in the same film in figure~\ref{Optical2}(a). The transition between non-metallic and metallic conductivity can be seen in Figure~\ref{Phase_D}(a) which shows the ratio of 300~K conductivity to 2~K conductivity for the series of SmN films. It is significant that the conductivity of SmN films which have exhibited superconductivity are found within this range, which we discuss in more details now.

\subsection{Preliminary phase diagram}

Figure~\ref{Phase_D}(b) shows a preliminary phase diagram depicting the paramagnetic, ferromagnetic and superconducting phases of SmN and the anti-ferromagnetic phase of Sm metal. The order parameter of choice is carrier concentration, for which we use the low temperature conductivity as a proxy. The filled circles on the plot show the transition between the paramagnetic and ferromagnetic phases. The temperature of this transition was estimated from the peak in the resistivity as discussed previously. The Curie temperature grows with increasing conductivity from $\sim$~20~K in the most undoped films to a maximum of $\sim$~30~K. This suggests an RKKY-enhanced exchange interaction. Exactly such an enhancement is seen also in GdN~\cite{Plank2011,Sharma2010} and DyN~\cite{Shaib2020}. Unlike those examples the end state of Sm metal is AFM and indeed the Curie temperature in Figure~\ref{Phase_D}(b) drops and finally looks to terminate near a nitrogen vacancy doping corresponding to a conductivity of $\sim$ 20~000~$\Omega^{-1}$cm$^{-1}$.

Although we have only unobserved clear superconductivity (via both a Meniser effect and a zero-resistance) in a few films~\cite{Anton2016} we regularly see the onset of a low temperature phase characterised by a resistance drop, which is often more clearly observed in the low temperature magneto-resistance. The onset of this phase is plotted as triangles in Figure~\ref{Phase_D} (the open triangles depict the films from ref.~\cite{Anton2016} which showed robust superconductivity). It is significant that the onset of the low temperature phase roughly tracks the onset of ferromagnetism, which is enhanced  in the more conductive films before a sharp drop. The crosses on the \textit{x}-axis of Figure~\ref{Phase_D} show films in which we did not observe a low temperature transition or onset about our minimum temperature of 1.9~K. It is interesting that almost all films in the conductivity range of 1~$\Omega^{-1}$cm$^{-1}$ to 1000~$\Omega^{-1}$cm$^{-1}$ show a low temperature transition, while most films above 1000~$\Omega^{-1}$cm$^{-1}$ do not. The scatter in the data ,particularly at high conductivity, highlight that $\sigma_{2K}$ is a rough proxy for the order parameter. We propose the carrier concentration, driven by nitrogen vacancy doping, is the most natural choice of order parameter and although the conductivity is proportional to this, there are other contributions for example the scattering time $\tau$ which may cause problems with this proxy, particularly in the most conductive films. 

The competition between the AFM and FM phases in the region above 1000~$\Omega^{-1}$cm$^{-1}$ is reminiscent of the competition between the Kondo and RKKY interactions proposed decades ago~\cite{Doniach1977}. Within that picture the ground state was RKKY-mediated AFM, changing to PM when the Kondo interaction dominates. In the present system Figure~\ref{Phase_D}(b) shows a falling Curie temperature above 1000~$\Omega^{-1}$cm$^{-1}$; the FM exchange coupling clearly weakens with the increasing dominance of the AFM interaction of Sm metal. It is significant that the superconducting transitions appears strongest just where the fall in Curie temperature shows the influence of FM / AFM competition.

The superconducting transition appears precisely where one would expect to observe unconventional superconductivity within the quantum critical point scenario. The spin-spin pairing mechanism is thought to be viable only near the critical point where long range magnetic order is suppressed~\cite{Mathur1998,Pfleiderer2009}. This study suggests a quantum critical point scenario scenario in the SmN-Sm phase diagram near a nitrogen vacency doping corresponding to 2000~$\Omega^{-1}$cm$^{-1}$. It is certainly suggestive that the most robust supercondtvity is observed in a region of Figure~\ref{Phase_D}(b) where the Curie temperature falls, indicating a weakening inter-ion pairing mechanism and strong quantum fluxuations. 

The present results which, along with previous experimental reports~\cite{Holmes-Hewett2018}, support electron transport in a completely spin-polarised defect band, suggest the origins of the superconducting state may indeed not be of the conventional \textit{s}-wave paring type. Any pair formed in the defect band of SmN must have S=1 and thus require an odd \textit{p}-wave orbital function.


\section{Conclusions}

We have undertaken experimental and computational studies on SmN films over a wide range of conductivities. Our findings indicating that SmN is insulating in the undoped form, exhibits a hopping type conductivity when dilutely doped, and becomes metallic like when doped with a significant concentration of nitrogen vacancies. Our comparison of XPS measurements to DOS calculations reinforces the use of the calculated Hubbard $U_f$ parameter used here and in previous studies. We have now systematically observed the in-gap defect states in SmN via optical spectroscopy and found that these correspond well with the calculated defect states. Both tracking to lower energy as the films are doped with nitrogen vacancies. Our present work suggests the localtion of a quantum critical point in the SmN-Sm phase diagram and makes a significant push towards the understanding of electrical transport in SmN which is required to begin understanding the unconventional superconducting state which co-exists with ferromagnetic order.

\section{ACKNOWLEDGMENTS}

This research was supported by the New Zealand Endeavour fund (Grant No. RTVU1810). The MacDiarmid Institute is supported under the New Zealand Centres of Research Excellence Programme.

\bibliography{master}

\end{document}